\def\year{2021}\relax
\documentclass[letterpaper]{article} 
\usepackage{aaai21}  
\usepackage{times}  
\usepackage{helvet} 
\usepackage{courier}  
\usepackage{pdfpages}
\usepackage[hyphens]{url}  
\usepackage{graphicx} 
\usepackage{graphicx}  
\usepackage[switch]{lineno}
\usepackage{amsmath}
\usepackage{bbm}
\usepackage{nameref}
\usepackage{etoolbox}
\usepackage{subfiles}

\makeatletter
\patchcmd{\maketitle}{\@copyrightspace}{}{}{}
\makeatother

\usepackage[square,sort,comma,numbers]{natbib}

\usepackage{url}            
\usepackage{booktabs}       
\usepackage{amsfonts}       
\usepackage{nicefrac}       
\usepackage{microtype}      

\usepackage{todonotes}
\usepackage[linesnumbered,ruled,vlined]{algorithm2e}
\SetCommentSty{mycommfont}

\SetKwInput{KwInput}{Input}                
\SetKwInput{KwOutput}{Output}              

\title{An Attention Mechanism using Multiple Knowledge Sources for COVID-19 Detection from CT Images}

\author{Duy M. H. Nguyen\textsuperscript{\rm 1},
Duy M. Nguyen\textsuperscript{\rm 2},
Huong Vu\textsuperscript{\rm 3},
Binh T. Nguyen\textsuperscript{\rm 4},
Fabrizio Nunnari\textsuperscript{\rm 1}
Daniel Sonntag\textsuperscript{\rm 1}\\
}

\author {
        Duy M. H. Nguyen,\textsuperscript{\rm 1, 5}
        Duy M. Nguyen, \textsuperscript{\rm 2}
        Huong Vu, \textsuperscript{\rm 3} 
        Binh T. Nguyen \textsuperscript{\rm 4}\\
        Fabrizio Nunnari, \textsuperscript{\rm 1}
        Daniel Sonntag\textsuperscript{\rm 1, 6}\\
}
 \affiliations {
     \textsuperscript{\rm 1} German Research Center for Artificial Intelligence, Saarbrücken, Germany\\
     \textsuperscript{\rm 2} School of Computing, Dublin City University, Ireland
  \\
  \textsuperscript{\rm 3} University of California, Berkeley
 \\
 \textsuperscript{\rm 4} VNUHCM-University of Science, Ho Chi Minh City, Vietnam\\
 \textsuperscript{\rm 5} Max Planck Institute for Informatics, Germany
 \\
  \textsuperscript{\rm 6} Oldenburg University, Germany 
  
 }

\begin{document}
\maketitle

\begin{abstract}
Until now, Coronavirus SARS-CoV-2 has caused more than 850,000 deaths and infected more than 27 million individuals in over 120 countries. Besides principal polymerase chain reaction (PCR) tests, automatically identifying positive samples based on computed tomography (CT) scans can present a promising option in the early diagnosis of COVID-19. Recently, there have been increasing efforts to utilize deep networks for COVID-19 diagnosis based on CT scans. While these approaches mostly focus on introducing novel architectures, transfer learning techniques or construction of large scale data, we propose a novel strategy to improve several performance baselines by leveraging multiple useful information sources relevant to doctors' judgments.  Specifically, infected regions and heat-map features extracted from learned networks are integrated with the global image via an attention mechanism during the learning process. This procedure makes our system more robust to noise and guides the network focusing on local lesion areas.  Extensive experiments illustrate the superior performance of our approach compared to recent baselines. Furthermore, our learned network guidance presents an explainable feature to doctors to understand the connection between input and output in a grey-box model.
\end{abstract}

\section{Introduction}
Coronavirus disease 2019 (COVID‑19) is a dangerous infectious disease caused by severe acute respiratory syndrome coronavirus 2 (SARS-CoV-2).  It was first recognized in December 2019 in Wuhan, Hubei, China, and continually spread to a global pandemic. According to statistics at Johns Hopkins University (JHU) \footnote{https://coronavirus.jhu.edu/map.html}, until the end of August 2020, COVID‑19 caused more than 850,000 deaths and infected more than 27 million individuals in over 120 countries.
Among the COVID-19 measures, the reverse- transcription-polymerase chain reaction (RT-PCR) is regularly used in the diagnosis and quantification of RNA virus due to its accuracy. However, this protocol requires functional equipment and strict requirements for testing environments, limiting the rapid of suspected subjects. Further, RT-PCR testing also is reported to suffer from high false-negative rates (\citeauthor{ai2020correlation} \citeyear{ai2020correlation}). For complementing RT-PCR methods, testings based on visual information as X-rays and computed tomography (CT) images are applied by doctors. They have demonstrated effectiveness in current diagnoses, including follow-up assessment and prediction of disease evolution (\citeauthor{rubin2020role} \citeyear{rubin2020role}). For instance, a hospital in China utilized chest CT for 1014 patients and achieved 0.97 of sensitivity, 0.25 of specificity compared to RT-PCR testing (\citeauthor{ai2020correlation} \citeyear{ai2020correlation}). (\citeauthor{fang2020sensitivity} \citeyear{fang2020sensitivity}) also showed evidences of abnormal CT compatible with an early screening of COVID-19. (\citeauthor{ng2020imaging} \citeyear{ng2020imaging}) conducted a study on patients at Shenzhen and Hong Kong and found that COVID-19's pulmonary manifestation is characterized by ground-glass opacification with occasional consolidation on CT.
Generally, these studies suggest that leveraging medical imaging may be valuable in the early diagnosis of COVID-19.

There have been several deep learning-based systems proposed to detect positive COVID-19 on both X-rays and CT imaging. Compared to X-rays, CT imaging is widely preferred due to its merit and multi-view of the lung. Furthermore, the typical signs of infection could be observed from CT slices, e.g., ground-glass opacity (GGO) or pulmonary consolidation in the late stage, which provide useful and important knowledge in competing against COVID-19. Recent studies focused on three main directions: introducing novel architectures, transfer learning methods, and building up a large scale for COVID-19.  For the first category, the novel networks are expected to discriminate precisely between COVID and non-COVID samples by learning robust features and less suffering with high variation in texture, size, and location of small infected regions. For example, (\citeauthor{wang2020deep} \citeyear{wang2020deep}) proposed a modified inception neural network (\citeauthor{szegedy2015going} \citeyear{szegedy2015going}) for classifying COVID-19 patients and normal controls by learning directly on the regions of interest, which are identified by radiologists based on the appearance of pneumonia attributes instead of training on entire CT images.  Although these methods could achieve promising performance, the limited samples could potentially simply over-fit when operating in real-world situations. Thus, in the second and third directions, researchers investigated several transfer learning strategies to alleviate data deficiency (\citeauthor{he2020sample} \citeyear{he2020sample}) and growing data sources to provide more large-sized datasets while satisfying privacy concerns and information blockade (\citeauthor{cohen2020covid} \citeyear{cohen2020covid}; \citeauthor{he2020sample} \citeyear{he2020sample}).

Unlike recent works, we aim to answer the question: ``\textit{how can we boost the performance of COVID-19 diagnosis algorithms by exploiting other source knowledge relevant to a radiologist's decision?}". Specifically, given a baseline network, we expect to improve its accuracy by incorporating properly two important knowledge sources: an infected and a heat-map region without modifying its architecture. In our settings, infected regions refer to positions of Pulmonary Consolidation Region (PCR) (as shown in figure 1 at the middle, green regions), a type of lung tissue filling with liquid instead of air; and Ground-Glass Opacity (GGO), an area of increased attenuation in the lung on CT images with preserved bronchial and vascular markings (as depicted in figure 1 at the middle, red regions).  By quantifying those regions, the radiologists can distinguish normal and infected COVID-19 tissues. While infected areas are based on medical knowledge, we refer to heat-map (as shown in figure 1 at the right-hand side) as a region extracted from a trained network, which allows us to understand transparently essential parts in the image directly impact the network decision. Our method motivates from the two following ideas. \textit{Firstly}, we would like to simulate how a radiologist can comprehensively consider both global, local information, and their prior knowledge to make final judgments by associating global images, infected regions, and heat-maps during the training process.  \textit{Secondly}, for avoiding network suffering by a significant level of noise outside the lesion area, an attention mechanism to supervise the network is necessarily such that it can take both lesion regions and global visual information into account for a final decision.

We introduce an attention mechanism to integrate all visual cues via a triplet stream network to realize those ideas. Our method can be highlighted in two attributes. First, it has two dedicated local branches to focus on local lesion regions, one for infected and another for heat-map areas. In this manner, the noise's influence in the non-disease areas and missing essential structures can be alleviated. Second, our principal branches, i.e., a global branch and two local branches, are connected by a fusion branch. While the local branches represent the attention mechanism, it may lead to information loss when the lesion areas are scattered in the whole image. Therefore, a global component is demanded to compensate for this error. We reveal that the global and local branches complement each other by the fusion branch, which shows better performance than the current state-of-the-art methods. 
\begin{figure}[bt!]
\begin{center}
	\includegraphics[width=0.15\textwidth]{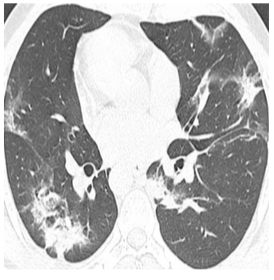}
	\includegraphics[width=0.15\textwidth]{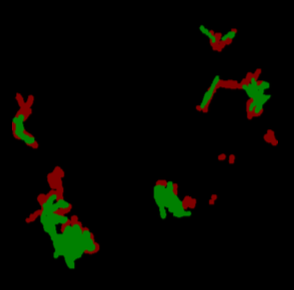}
		\includegraphics[width=0.15\textwidth]{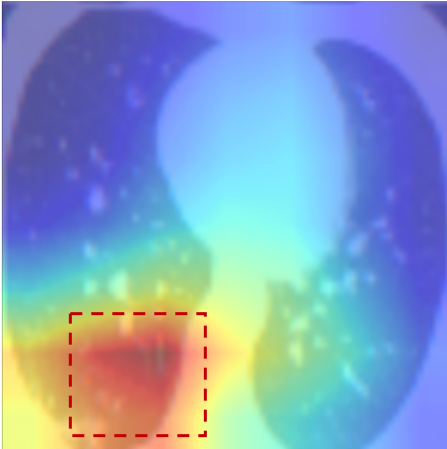}
	\caption{\small Left: the picture of a COVID-19 case. Middle: red and green labels indicate the Ground-Glass Opacity (GGO) and Pulmonary Consolidation regions (\citeauthor{fan2020inf} \citeyear{fan2020inf}). Right: heat-map region extracted from trained network.}
\end{center}
	\label{fig:lesion-examples}
\end{figure}

In summary, we make two following contributions:
\begin{itemize}
    \item We provide a new procedure to advance baselines on COVID-19 diagnosis without modifying the network's structures by integrating knowledge relevant to radiologists' judgment as examining a suspected patient. Extensive experiments demonstrate that the proposed method can boost several cutting-edge models' performance, yielding a new state-of-the-art achievement. 
    
    \item We show the transparency of learned features by embedding the last layer's output vector in the fusion branch to smaller space and visualizing in a 3-D dimension (as shown in figure 3). Interestingly, we found a strong connection between learned features and network decisions as mapping of activation heat-map and infected regions. Such property is a critical point for clinicians as end-users, as they can interpret how networks create a result given input features in a grey-box rather than a black-box algorithm.
\end{itemize}


\section{Related Works}

\label{subsec:diagnosis}
In a global effort against COVID-19, the computer vision community pays attention  
on constructing efficient deep learning approaches to perform screening of COVID-19 in CT scans. (\citeauthor{zheng2020deep} \citeyear{zheng2020deep}) pioneered in introducing a novel 3D-deep network  (DeCoVNet) composed from pre-trained U-net (\citeauthor{ronneberger2015u} \citeyear{ronneberger2015u}) and two 3D residual blocks. To reduce annotating costs, the authors employed  weakly-supervised based computer-aided COVID-19 detection with a large number of CT volumes from the frontline hospital. Other methods also applied 3D deep networks for CT images can be found in (\citeauthor{gozes2020rapid} \citeyear{gozes2020rapid}; \citeauthor{li2020artificial} \citeyear{li2020artificial}). Recently, there are also two other state of the arts from works of (\citeauthor{saeedi2020novel} \citeyear{saeedi2020novel}) and (\citeauthor{mobiny2020radiologist} \citeyear{mobiny2020radiologist}), which trained directly on 2D images on a dataset collected from (\citeauthor{he2020sample} \citeyear{he2020sample}) with $746$ CT samples. While (\citeauthor{saeedi2020novel} \citeyear{saeedi2020novel}) developed a novel method by combining several pre-trained networks on ImageNet with regularization of support vector machine, (\citeauthor{mobiny2020radiologist} \citeyear{mobiny2020radiologist}) proposed a novel network, namely DECAPS, by leveraging the strength of Capsule Networks with several architecture to boost classification accuracies. In other 
trends, (\citeauthor{song2020deep} \citeyear{song2020deep}) developed CT diagnosis to support clinicians to identify patients with COVID-19 based on the presence of Pneumonia feature.

To mitigate data deficiency, (\citeauthor{chen2020simple} 
\citeyear{chen2020simple}) build a publicly-available dataset containing hundreds of CT scans that are positive for COVID-19 and introducing a novelty sample-efficient method based on both pre-trained ImageNet (\citeauthor{deng2009imagenet} \citeyear{deng2009imagenet}) and self-supervised learning. In the same effort, (\citeauthor{cohen2020covid} \citeyear{cohen2020covid}) also contributes open image data collection, created by assembling medical images from websites and publications. While recent networks only tackle in a sole target, e.g., only diagnosis or compute infected regions.  In contrast,  we bring those components into a single system by fusing straight infected areas and global images throughout the learning-network procedure so that these sources can support each other to make our model more robust and efficient.

\begin{figure*}[hbt!]
    \centering
    \includegraphics[width=0.69\textwidth]{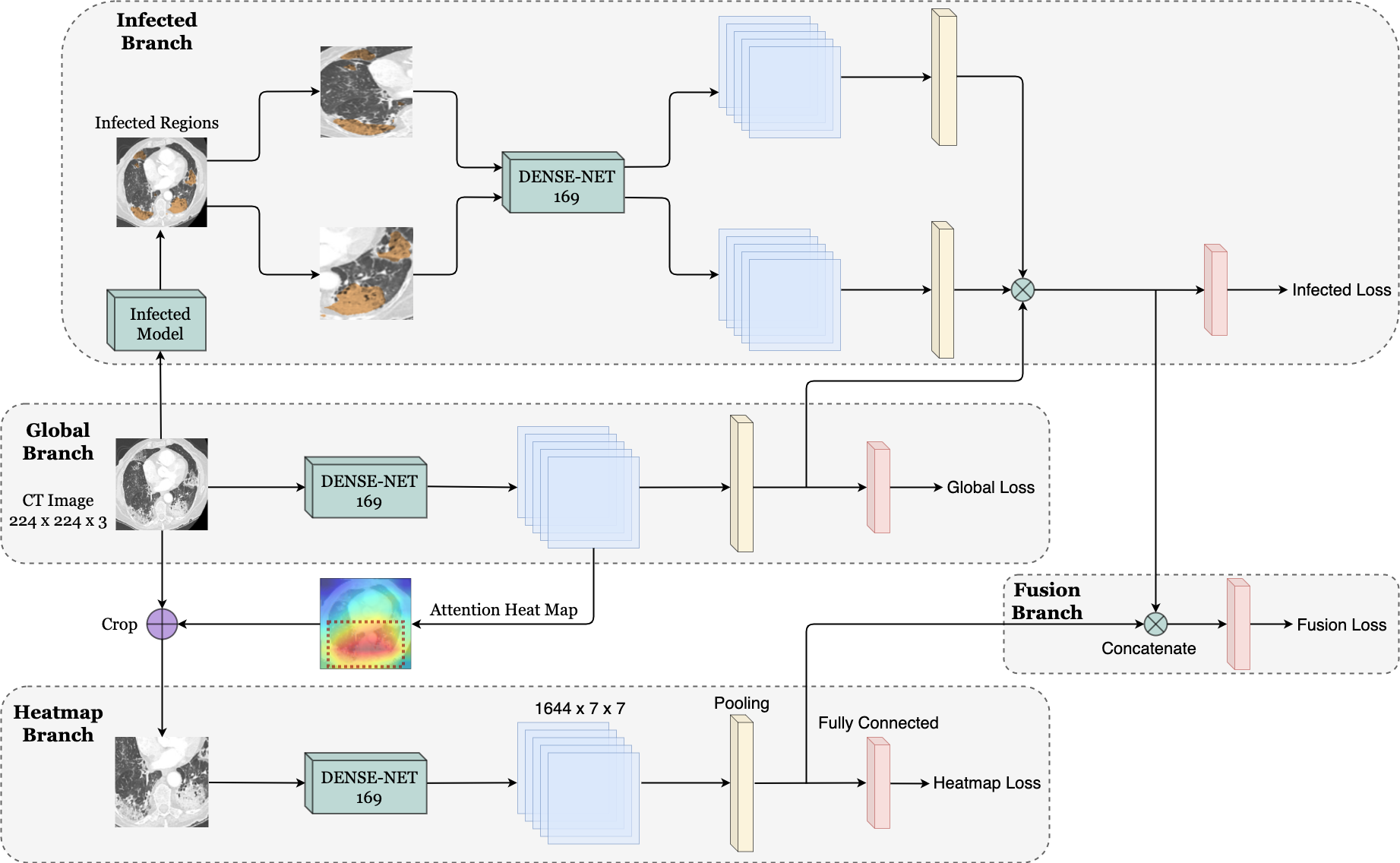}
    \caption{\small Our proposed attention mechanism given a specific backbone network to leverage efficiently three knowledge sources: infected regions (top branch), global image (middle branch) and learned heat-maps (bottom branch). For all branches, we utilize a binary cross entropy loss function during the training process. The backbone network (DenseNet-169 in this figure) can be replaced by arbitrary networks in general case.}
    \label{fig:overview}
\end{figure*}


\section{Methodology}

\label{sec:methods}
\subsection{Fusion with Multiple Knowledge}
\label{sec:fusion}
\subsubsection{Infected Branch}
In (\citeauthor{fan2020inf} \citeyear{fan2020inf}), authors developed methods to identify lung areas that are infected by ground-class opacity and consolidation by presenting a novel architecture, namely \textit{Inf-Net}. Given the fact that
there is a strong correlation between the diagnosis of COVID-19 and ground-class opacity presented in lung CT scans. Therefore, we adopt the Semi-Infected-Net method from (\citeauthor{fan2020inf} \citeyear{fan2020inf}) to localize lung areas suffered by ground-class opacity and consolidation on our CT images. In particular, we expect using this quantification to reduce focused regions of our model to important positions, thus making the system learn efficiently. 


Following approach based on semi-supervised data in (\citeauthor{fan2020inf} \citeyear{fan2020inf}), we extend it in the diagnosis task by first training the \textit{Inf-Net} on D1 dataset (please see Section \nameref{sub:data} for further reference). Then, we use this model to obtain pseudo label segmentation masks for $100$ randomly chosen CT images from D2 and D3 datasets. After that, we combine the newly predicted masks with D1 as a new training set and re-train our model. The re-trained model will continue to be used for segmenting other $100$ ones randomly chosen from the remaining D2 and D3. Then, we repeated this data combining step. The cycle continues until all images from D2 and D3 have a segmentation mask. We summarize the whole procedure in algorithm 1.
\begin{algorithm}[!hbt]
\DontPrintSemicolon
  
  \KwInput{$D_{\text{train}}$ = D1 with segmentation masks and \\ \ \ \ \ \ \  $\ \ \ \ \ \ D_{\text{test}}$ = D2 $\cup$ D3 without masks.}
  \KwOutput{Trained Infected Net model, $M$}
  \textbf{Set} $D_{\text{train}}$ = D1; $D_{\text{test}}$ = D2 $\cup$ D3; $D_{\text{subtest}}$ = NULL\\
  \While{len($D_{\text{test}}) > 0$}
  {
   Train $M$\\
   \If{len($D_{\text{test}} > 100$)}
        {
            $D_{\text{subtest}}$ = random $(\ D_{\text{test}}\backslash D_{\text{subtest}}, \text{k = 100})$\\
            $D_{\text{train}} = D_{\text{train}} \cup  M(D_{\text{subtest}})$\\
            $D_{\text{test}} = D_{\text{test}}\backslash D_{\text{subtest}}$
        }
    \Else
        {
            $D_{\text{subtest}} = D_{\text{test}}$\\
            $M(D_{\text{subtest}})$\\
            $D_{\text{test}} = D_{\text{test}}\backslash D_{\text{subtest}}$
        }
    }
\caption{Training Semi-supervised Infected Net}
\label{al:semi-inf}
\end{algorithm}

%
%
%

\subsubsection{Heat-map Branch} \label{sec:heat-mapbranch}
Besides the whole original scans of CT images, we wanted our proposed network to pay more attention to injured regions within each image by building a heat-map branch, which was a separate traditional classification structure as DenseNet169 (\citeauthor{huang2017densely} \citeyear{huang2017densely}) or ResNet50 backbone (\citeauthor{he2016deep} \citeyear{he2016deep}). This additional model was expected to learn the discriminative information from a specific CT scan area instead of the entire image, hence alleviating noise problems. 

A lesion region of a CT scan, which could be considered as an attention heat-map, was extracted from the last convolution layer's output before computing the global pooling layer of the backbone (DenseNet169 or ResNet50) in the main branch. In particular, with an input CT image, let $f_k(x,y)$ be the activation unit in the channel $k$ at the spatial $(x, y)$ of the last CNN layer, in which $k \in \{1,2,...,K\}$ and $K=1644$ for DenseNet169 or $K=2048$ for ResNet50 as a backbone. Its attention heat-map, $H$, is created by normalizing across $k$ channels of the activation output by using Eq.~\ref{eq:heat-map}.

\begin{equation}
\label{eq:heat-map}
H(x,y) = \frac{\sum_{k} f_k(x,y) - \min(\sum_{k} f_k)}{\max(\sum_{k} f_k)}
\end{equation}
We then binarized $H$ to get the mask $B$ of the suspected region in Eq.~\ref{eq:mask}, where $\tau$ is a tuning parameter whose smaller value produces a larger mask, and vice versa.
\begin{equation}
\small
\label{eq:mask}
  B =
    \begin{cases}
      1, & \text{if $H(x,y) > \tau$}\\
      0, & \text{otherwise}\\
    \end{cases}       
\end{equation}
We then extracted a maximum connected region in $B$ and mapped with the original CT scan to get our local branch's final input. One can see a typical example of the heat-map area in figure 1 on the right-hand side. Given this output and coupling with an infected model $M$ obtaining from algorithm 1, we now have enough input to start training the proposed model.
\subsection{Network Design and Implementation}
\label{sub:network_design}
\subsubsection{Multi-Stream network}
Our method's architecture can be illustrated in figure \ref{fig:overview}, with DenseNet169 as an example of the baseline model. It has three principal branches, i.e., the global and two local branches for attention lesion structures, followed by a fusion branch at the end. Both the global and local branches play roles as classification networks that decide whether the COVID-19 is present. Given a CT image, the parameters of \textit{Global Branch} are first fine-tuned by loading either pre-trained ImageNet or Self-transfer learning tactics as in (\citeauthor{he2020sample} \citeyear{he2020sample}), and continue to train on global images. Then, heat-map regions from the global image extracted using 
equations (\ref{eq:heat-map}) and (\ref{eq:mask}) are utilized as an input to train on \textit{heat-map Branch}.  In the next step, input images at the \textit{Global Branch} are fed into Infected-Model $M$, which is derived after completing the training procedure in algorithm \ref{al:semi-inf}, to produce infected regions.
Because these lesion regions are relatively small, disconnected, and distributed on the whole image, we find bounding boxes to localize those positions and divide it into two sub-regions: left infected and right infected photos. Those images can be fed into a separate backbone network to output two pooling layers and then concatenating with pooling features from the global branch to train for \textit{Infected Branch}. It is essential to notice that concatenating output features from \textit{Infected Branch} with global features is necessary since, in several cases, e.g., in healthy patients, we could not obtain infected regions. Finally, the \textit{Fusion Branch} can be learned by merging all pooling layers from both global and two local branches. 

To be tighter, we assume that each pooling layer is followed by a fully connected layer $FC$ with $C-$ dimensional for all branches and a sigmoid layer is added to normalize the output vector. Denoting $(I_g, W_g,\, p_{g}(c|I_{g})),\, (I_h,\, W_{h},\, p_{h}(c|I_{g}, I_{h})),$ and $ \, (I_{in}, \,W_{in},\, p_{in}(c|I_{g}, I_{in}))$ as pairs of images, parameters and probability scores belong to the $c$-th class, $\small c \in  \{1, 2\,..., C\}$ at $FC$ layer for global, heat-map and infected branches, respectively. For each fushion branch, we also denote $(Pool_{k}, \, W_{f}, \, p_{f}(c|(I_g, I_h, \, I_{in}))$ as a pair of output 
feature at pooling layer in branch $k\  (k \in \{g,\,h,\,in\})$, parameter and probability scores belong to the $c$-th class of the fusion branch. Then, parameters $W_g, W_h$, and $W_{in}$ are optimized by minimizing the binary cross-entropy loss as follows:
\begin{equation}
\label{eq:loss}
    L(W_{i}) = -\frac{1}{C}\sum_{c=1}^{C}l_c\log(\tilde{p}_{i}(c)) + (1 - l_c)\log(1-\tilde{p}_{i}(c)),
\end{equation}
where $l_c$ is the ground-truth label of the $c$-th class, $C$ is the total of classes, and $\tilde{p}_{i}(c)$ is the normalized output network at branch $i$ $(i \in \{g,\,h,\,in\})$, which can be computed by:
\begin{equation}
\label{eq:softmax}
    \tilde{p_i}(c) = 1/(1 + \exp(-p_{i}(c|I_g, I_h, I_{in})
\end{equation}
in which 
\begin{equation}
    p_i(c|I_g, I_h, I_{in}) =
\left\{\begin{matrix}
p_{g}(c|I_g)\ \ \textrm{if}\ \  i=g\\ 
p_{h}(c|I_g, I_h) \ \  \textrm{if}\ \  i=h\\
p_{in}(c|I_g, I_{in})\ \  \textrm{if}\ \  i=in
\\ 
\end{matrix}\right.
\end{equation}
\\
For the fusion branch, we have to compute the pooling fusion 
$Pool_{f}$ by merging all pooling values in all branches: $Pool_{f} = \left[Pool_{g},\ Pool_{h}, Pool_{in}\right]$. After that, we evaluate $p_{f}(c|(I_g, I_h, I_{in})$ by multiplying $Pool_f$ with weights at $FC$ layer. Finally, $W_{f}$ can be learned by minimizing equation (\ref{eq:loss}) with formula (\ref{eq:softmax}).

\subsubsection{Training Strategy}
Due to the limited amount of COVID-19 CT scans, it is not suitable to simultaneously train entire branches. We thus proposed a strategy that trains each part sequentially to reduce the number of parameters being trained at once. As a branch finished its training stage, its weights would be used to initialize the next branches. Our training protocol can be divided into three stages, as follows:

\textbf{\textit{Stage I}}: We firstly trained and fine-tuned the global branch, which used architectures from an arbitrary network such as DenseNet169 or ResNet50. The weight initialization could be done by loading pre-trained ImageNet or Self-Transfer learning method (\citeauthor{he2020sample} \citeyear{he2020sample}).

\textbf{\textit{Stage II}}: Based on the converged global model, we then created attention heat-map images to have the input for the heat-map branch, which was fine-tuned based on the hyper-parameter $\tau$ as described in section \textit{\nameref{sec:heat-mapbranch}}. Simultaneously, we could also train the infected branch independently with the heat-map branch using the pooling features produced by the global model, as illustrated in figure 2. The weights of the global model were kept intact during this phrase.

\textbf{\textit{Stage III}}: Once the infected branch and the heat-map branch were fine-tuned, we concatenated their pooling features and trained our final fusion branch with a fully connected layer for the classification. All weights of other branches were still kept frozen while we trained this branch.

The overall training procedure was summarized in algorithm 2. Different training configurations might affect the performance of our system. Therefore, we analyzed this impact from variation training protocol in experiment results.
\begin{algorithm}[!hbtp]
\DontPrintSemicolon
  \KwInput{Input image $I_g$, Label vector $L$, Threshold $\tau$}
  \KwOutput{Probability score $p_{f}(c|I_g, I_h, I_{in})$}
  Learning $W_g$ with I, computing $\tilde{p_g}(c|I_g)$, optimizing by Eq. \ref{eq:loss} (\textbf{Stage I}); \\
  
  Finding attention heat-map and its mapped image $I_h$ of $I_g$ by Eq. \ref{eq:mask} and Eq. \ref{eq:heat-map}. \\
  
  Learning $W_h$ with $I_h$, computing $\tilde{p_h}(c|I_g, I_h)$, optimizing by Eq. \ref{eq:loss} (\textbf{Stage II}); \\
  
  Finding infected images $I_{in}$ of $I_g$ by using infected model $M$; \\
  
  Learning $W_{in}$ with $I_{in}$, computing $\tilde{p_{in}}(c|I_g, I_{in})$, optimizing by Eq. \ref{eq:loss} (\textbf{Stage II}); \\
  
  Computing the concatenated $Pool_f$, learning $W_f$, computing $p_f(c|I_g, I_h, I_{in})$, optimizing by Eq. \ref{eq:loss} (\textbf{Stage III}).
\caption{Training our proposed system}
\label{al:system}
\end{algorithm}

\section{Experiment and Results}
This section presents our settings, chosen datasets, and the corresponding performance of different methods.
\subsection{Data}
\label{sub:data}
In our research, we use three sets of data.
\begin{itemize}
    \item \textit{D1. COVID-19 CT Segmentation from “COVID-19 CT segmentation dataset”}\footnote{https://medicalsegmentation.
com/covid19/}.

    This collection contains 100 axial CT images of more than 40 COVID-19 patients with labeled lung area and associating with ground-class opacity, consolidation, and pleural effusion \ref{fig:lesion-examples}.
    \item \textit{D2. COVID-19 CT Collection from (\citeauthor{fan2020inf} \citeyear{fan2020inf})}.
    
    This dataset includes 1600 CT slices, extracted from 20 CT volumes of different COVID-19 patients. Since these images are extracted from CT volumes, they do not have segmentation masks.
    \item \textit{D3. Sample-Efficient COVID-19 CT Scans from (\citeauthor{he2020sample} \citeyear{he2020sample})}.
    
    This data comprises 349 positive CT images form 216 COVID-19 patients and 397 negative CT images selected from the PubMed Central \footnote{ https://www.ncbi.nlm.nih.gov/pmc/} and publicly-open online medical image database \footnote{https://medpix.nlm.nih.gov/home}. D3 also does not have segmentation masks; only COVID-19 positive/negative labels are involved.
\end{itemize}

For all experiments, we exploited all datasets for training the Infected-Net model while detection performance was evaluated on the D3 dataset.  

\label{sec:data}
\subsection{Settings}
We implemented several experiments on a TITAN RTX GPU with the Pytorch framework.
The optimization used SGD with a learning rate of 0.01 and is divided by ten after $30$ epochs. We configured a weight decay of 0.0001 and a momentum of 0.9. For all baseline networks, we used a batch size of $32$ and training for each branch $50$ epochs with input size $224 \times 224$. The best model is chosen based on early stopping on validation sets. We optimized hyper-parameters $\tau$ by grid searching with 0.75, which yielded the best performance on the validation set.
\begin{table}[!hbt]
  \centering
    \scalebox{0.85}{
  \begin{tabular}{lccc}
    \toprule
    \cmidrule(r){1-2}
    \textbf{Method}     & \textbf{Accuracy} & $\mathbf{F_1}$   & \textbf{AUC}  \\
    \midrule
    \textbf{ResNet50} $\mathbf{^{(1)}}$ \textit{(ImgNet, Global)}	 & 0.803	 &  0.807     & 0.884\\
    \textbf{DenseNet169} $\mathbf{^{(1)}}$	\textit{(ImgNet, Global)} & 0.832	 &  0.809        &0.868\\
    \midrule
  \textbf{ResNet50} $\mathbf{^{(1)}}$ + \textit{Our Infected }	 & 0.831 	 & 0.815      &0.897\\
  \textbf{ResNet50} $\mathbf{^{(1)}}$ + \textit{Our heat-map} &  0.824& \textcolor{blue}{0.832} & 0.884\\
  \textbf{ResNet50} $\mathbf{^{(1)}}$ + \textit{Our Fusion}& \textcolor{blue}{0.843}  & 0.822 & \textcolor{blue}{0.919}\\
      \midrule
 \textbf{DenseNet169} $\mathbf{^{(1)}}$ + \textit{Our Infected}	 & 0.861 	 & 0.834     & 0.911 \\
   \textbf{DenseNet169} $\mathbf{^{(1)}}$ + \textit{Our heat-map}& 0.855 & 0.825 & 0.892\\
  \textbf{DenseNet169} $\mathbf{^{(1)}}$ + \textit{Our Fusion} & \textcolor{red}{0.875}  &\textcolor{red}{0.845}  & \textcolor{red}{0.927}\\
    \bottomrule
  \end{tabular}}
   \caption{\small Performance of two best  architectures on D3 dataset using \textbf{pre-trained ImageNet} with only used global images (ResNet50 $^{(1)}$, DenseNet169 $^{(1)}$) and obtained results by utilizing our strategy. Blue and Red colour are best values for ResNet50 and DenseNet169 correspondingly.}
  \label{tab:imagenet}
    \vspace{-0.2in}
\end{table}
\begin{table}[!hbt]
  \centering
  \scalebox{0.85}{
  \begin{tabular}{lccc}
    \toprule
    \cmidrule(r){1-2}
    \textbf{Method}     & \textbf{Accuracy} & $\mathbf{F_1}$   & \textbf{AUC}  \\
    \midrule
    \textbf{ResNet50} $\mathbf{^{(2)}}$ \textit{(Self-trans , Global)}	 & 0.841	 &  0.834     & 0.911\\
   \textbf{DenseNet169} $\mathbf{^{(2)}}$ \textit{(Self-trans , Global)}	 & 0.863	 &  0.852       &0.949\\
       \midrule
 \textbf{ResNet50} $\mathbf{^{(2)}}$ +  \textit{Our  Infected}	 & 0.842	 &   0.833   & 0.918 \\
 \textbf{ResNet50} $\mathbf{^{(2)}}$ +  \textit{Our heat-map} & \textcolor{blue}{0.879}  & 0.848 & 0.924\\
 \textbf{ResNet50} $\mathbf{^{(2)}}$ + \textit{Our Fusion}  & 0.861  & \textcolor{blue}{0.870} & \textcolor{blue}{0.927}\\
 \midrule
 \textbf{DenseNet169}  $\mathbf{^{(2)}}$ + \textit{Our Infected}	 & 	0.853 &   0.849   &0.948\\
 \textbf{DenseNet169} $\mathbf{^{(2)}}$ +   \textit{Our heat-map} &0.870  &0.837  & 0.954\\
  \textbf{DenseNet169} $\mathbf{^{(2)}}$ +  \textit{Our Fusion} & \textcolor{red}{0.882}  & \textcolor{red}{0.853} & \textcolor{red}{0.964}\\
    \bottomrule
  \end{tabular}}
  \caption{\small Performance of two best architectures on D3 dataset using \textbf{Self-trans} with only used global images ((ResNet50 $^{(2)}$, DenseNet169 $^{(2)}$)) and obtained results by utilizing our strategy. Blue and Red colour are best values for ResNet50 and DenseNet169 correspondingly.}
  \label{tab:selftrans}
  \vspace{-0.2in}
\end{table}

\subsection{Evaluations}
In this section, we evaluated our attention mechanism with different settings, such as semi-supervised procedure (algorithm \ref{al:semi-inf}) and training strategies (algorithm \ref{al:system}) on the D3 dataset. We also illustrated how our framework \textit{allowing to boost the performance of several baseline networks without modifying their architectures}. 

\begin{table*}[!tb]
  \centering
    \scalebox{0.85}{
  \begin{tabular}{lccc}
    \toprule
    \cmidrule(r){1-2}
    \textbf{Method}     & \textbf{Accuracy} & $\mathbf{F_1}$   & \textbf{AUC}  \\
    \midrule
    \textbf{Saeedi et al. 2020} 	 & 0.906 ($\pm 0.05$)	 &  0.901 ($\pm 0.05$)    & 0.951 ($\pm 0.03$)\\
    \textbf{Saeedi et al. 2020}  + \textit{Our Fusion w/out Semi-S} & 0.913 ($\pm 0.03$) 	 &  \textcolor{blue}{0.926 ($\pm 0.03$)}        & 0.960 ($\pm 0.03$) \\
    \textbf{Saeedi et al. 2020}  + \textit{Our Fully Fusion} & \textcolor{blue}{0.925 ($\pm 0.03$)} 	 &  0.924   ($\pm 0.03$)     &\textcolor{blue}{0.967 ($\pm 0.03$)} \\
    \midrule
  \textbf{Mobiny et al. 2020}   $\mathbf{^{(1)}}$	& 0.832 ($\pm 0.03$)	 & 0.837 ($\pm 0.03$)     &0.927 ($\pm 0.02$)\\
  \textbf{Mobiny et al. 2020}   $\mathbf{^{(1)}}$+ \textit{Our Fusion w/out Semi-S} &0.856 ($\pm 0.03$)  & 0.864 ($\pm 0.03$) & \textcolor{red}{0.950 ($\pm 0.02$)}\\
    \textbf{Mobiny et al. 2020}   $\mathbf{^{(1)}}$+ \textit{Our Fully Fusion} & \textcolor{red}{0.868 ($\pm 0.03$)}  & \textcolor{red}{0.872 ($\pm 0.03$)} & 0.947 ($\pm 0.02$)\\
      \midrule
  \textbf{Mobiny et al. 2020}  $\mathbf{^{(2)}}$	 & 0.876 ($\pm 0.01$)	 & 0.871 ($\pm 0.02$)     & 0.961 ($\pm 0.01$) \\
\textbf{Mobiny et al. 2020}   $\mathbf{^{(2)}}$+ \textit{Our Fusion w/out Semi-S} &0.885 ($\pm 0.01$)  & 0.884 ($\pm 0.02$)& 0.983 ($\pm 0.01$) \\
  \textbf{Mobiny et al. 2020}   $\mathbf{^{(2)}}$+ \textit{Our Fully Fusion} & \textbf{0.896} ($\mathbf{\pm 0.01}$) &\textbf{0.889} ($\mathbf{\pm 0.01}$)  &\textbf{0.986} \textbf{($\mathbf{\pm 0.01}$)}\\
    \bottomrule
  \end{tabular}}
   \caption{\small Performance of other state of the arts from
   (\citeauthor{saeedi2020novel} \citeyear{saeedi2020novel}) (the first row) and (\citeauthor{mobiny2020radiologist} \citeyear{mobiny2020radiologist}) (two options are represented by the fourth and seventh row)
 with only used global images  and obtained results by utilizing our strategy with multiple knowledge sources. Blue, red and bold colors represent the best values in each method.}
  \label{tab:otherbaseline}
  \vspace{-0.1in}
\end{table*}

\subsubsection{Improving on Standard Backbone Networks}
We first examined our approach's effectiveness on commonly deep networks like VGG-16, ResNet-18, ResNet-50, DenseNet-169, and EfficientNet-b0. Based on summarized results from (\citeauthor{he2020sample} \citeyear{he2020sample}), we picked two top networks that achieved the highest results on the D3 dataset and configuring them in our framework under two settings: initializing weights from pre-trained ImageNet or self-transfer techniques proposed in (\citeauthor{he2020sample} \citeyear{he2020sample}). We first used only global images for cases and then added one by one other option as heat-map, Infected, and Fusion branch to capture each component's benefits. Furthermore, the proposed training strategy (algorithm \ref{al:system}) and semi-supervised techniques (algorithm \ref{al:semi-inf}) were also involved.
\paragraph{\textit{Fusion Branch:}}
From both table 1 and table 2, it is clear that our fusion mechanism with ResNet50 and DenseNet169 has significantly improved performance compared to the default settings (only used global images) for all categories: pre-trained ImageNet and Self-Transfer Learning. By employing pre-trained ImageNet with ResNet50 backbone, our fusion method increases the accuracy from $80.3\%$ to $84.3\%$, which is slightly better than this network's accuracy using Self-Transfer Learning ($84.1\%$). Similarly, for DenseNet169 with pre-trained ImageNet, our fusion method can improve the performance from $83.2\%$ to $87.5\%$ in terms of accuracy. This accuracy once again is better than the option using Self-Transfer Learning ($86.3\%$). Our fusion method's outstanding performance is also consistent for two other metrics as AUC and $F_1$. With Self-Transfer (table 2), we continue boosting performance for both ResNet50 and DenseNet169, especially with the DenseNet169, a new milestone with $88.2\%$ and $96.4\%$ in Accuracy and AUC metrics is achieved,  which is higher $2\%$ compared to the original one.
\vspace{-0.1in}
\paragraph{\textit{Mixing Global and Local Branch:}} Using Infected information or heat-map with the baseline can boost the result from $3$ - $4\%$. For instance,  the Global-Infected structure for ResNet50 with pre-trained ImageNet (table 1) improves the exactness from $80.3\%$ to $83.1\%$. The Global-heat-map increases  ResNet50 with Self-Trans initialization (table 2) from $84.1\%$ to $87.9\%$. However, overall,  there is no pattern to conclude if either the Infected or heat-map branch outperforms the other. Furthermore, in most cases, the best values across metrics are obtained using the Fusion branch. This evidence demonstrates that using more relative information, more accurate predictions the model could make. 
\vspace{-0.1in}
\paragraph{\textit{Peformance of Training Strategies:}}
To validate the impact of the proposed training strategy (algorithm \ref{al:system}), we tested with various settings, for example, train all branches together, train global, heat-map, and infected together. These results can be found in table 1 appendix.  In general, training for each component sequentially is the most efficient case. This phenomenon might be due to the lack of the data as training the whole complex network simultaneously with the limited resources was not a suitable schema. Thus, training each branch independently and then fusing them can be the right choice in such situations. 


\subsubsection{Improving on State of The Arts}
In this experiment, we aim to further evaluate the proposed method's effectiveness by integrating the current state of the art methods on the D3 dataset. This includes three methods, one from (\citeauthor{saeedi2020novel} \citeyear{saeedi2020novel}) and two others from (\citeauthor{mobiny2020radiologist} \citeyear{mobiny2020radiologist}). Specifically, we used trained models following descriptions of authors and available code to plug in our framework. The experimental results in table 3 were calculated as the experimental design of each paper, for instance, ten-fold cross-validation in (\citeauthor{saeedi2020novel} \citeyear{saeedi2020novel}) and average of the best five trained model checkpoints in (\citeauthor{mobiny2020radiologist} \citeyear{mobiny2020radiologist}). Furthermore, the contribution of the semi-supervised strategy was also evaluated in various metrics for each method. 
\vspace{-0.1in}
\paragraph{\textit{Performance of Fully Settings:}}
"Fully settings" refers to utilizing the training method as in algorithm \ref{al:system} with fusing all branches. Interestingly, our attention method continues improving for all of these states of the art methods, resulting in obtaining a new benchmark \textit{without modifying available architectures.} Specifically, we boosted approximately $2\%$ for the method in (\citeauthor{saeedi2020novel} \citeyear{saeedi2020novel}) (from $90.6\%$ to $92.5\%$) and second option in (\citeauthor{mobiny2020radiologist} \citeyear{mobiny2020radiologist}) (from $87.6\%$ to $89.6\%$) in terms of accuracy metric. It is even better for the first option of (\citeauthor{mobiny2020radiologist} \citeyear{mobiny2020radiologist}) with an improvement up to $3.6\%$ (from $83.2\%$ to $86.8\%$). This benefit was also attained for other metrics as F1 and AUC. In short, this evidence once again confirmed the proposed method's effectiveness. A better result can be obtained by just using an available trained model and inserting it into our framework. In other words, our attention mechanism can be played as an "enhancing technique" in which the \textit{performance of a specific method can be improved by integrating properly multiple useful information relevant to doctors' judgments by our framework.}

\begin{figure*}[!hbtp]
    \centering
    \includegraphics[width=0.57\textwidth]{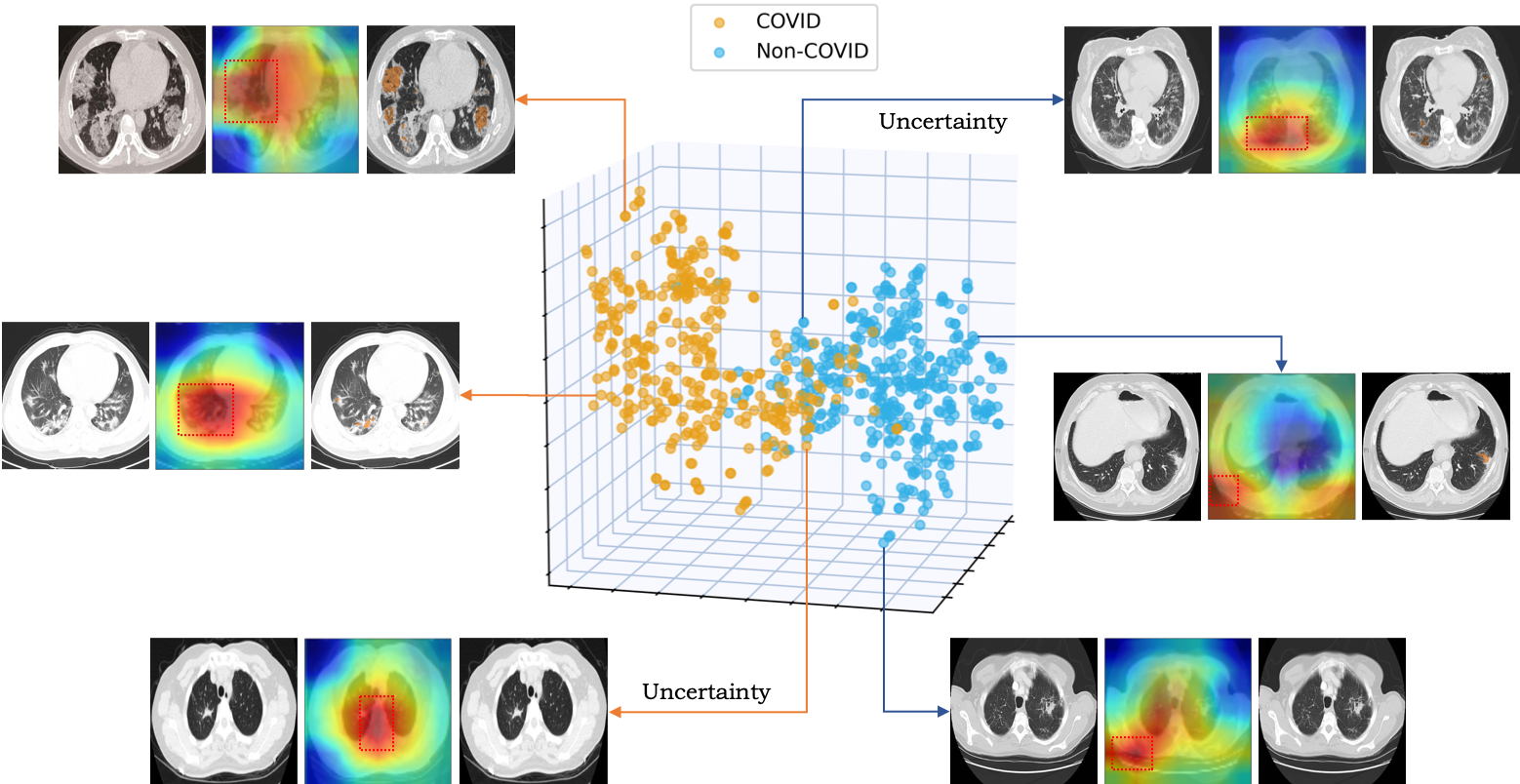}
    \caption{\small Interpreting learned features by t-SNE with the final layers of the fusion branch. Each point is presented together with its original scan, class activation map (CAM) representation, and infected regions (left to right order). For Covid and Non-Covid cases whose distance is far away from a decision margin, important heat-map regions (inside the rectangle) locate inside/outside the lung regions (zooming for better visualization). For points locating near the boundary margin, the heat-map area overlaps both the lung and non-lung area, which indicates for uncertainty property of the network's decision.}
    \label{fig:tsne_visualize}
\vspace{-0.1in}
\end{figure*}
\paragraph{\textit{Performance of Semi-Supervised:}}
The advantages of applying semi-supervised in final performance are also presented in table 3. 
Accordingly, without using semi-supervised tactics contributes a smaller improvement to the arts in most cases. Excepting the cases of (\citeauthor{saeedi2020novel} \citeyear{saeedi2020novel}) with F1 and the first version of (\citeauthor{mobiny2020radiologist} \citeyear{mobiny2020radiologist}) with AUC metric, without semi-supervised is better, however the difference is not significantly compared to fully settings.

\subsection{Interpretable Learned Features}
Besides high performance, an ideal algorithm should be explainable to doctors about its connection between learned features and the final network decision. Such property is critical, especially in medical applications; thereby the reliability is the most concerning factor. Furthermore, in our experiment, given that the D3 dataset only contains two classes Covid or Non-Covid, understanding how the model makes a decision is even more critical because it allows doctors to believe or not predict the trained model.
To answer this question, we interpret our learned features by generating the class activation map (CAM) (\citeauthor{zhou2015cam} \citeyear{zhou2015cam}) of the fusion branch and applied t-Distributed Stochastic Neighbor Embedding (t-SNE) (\citeauthor{maaten2008visualizing} \citeyear{maaten2008visualizing}) method for visualization by compressing $1644$-dimensional features (DenseNet169 case with Self-Trans) into a 3D space. Figure 3 depicts the pooling features' distribution on testing images of the D3 dataset using t-SNE and CAM representations. Furthermore, infected regions were also shown with their corresponding CT images. 

By considering CAM color and its corresponding labels, figure 3 indicated that for data points whose positions are far from the margin decision (both left and right), our system could focus precisely regions within the lesion lung area for positive scans and vice versa, the red heat-map parts locate outside the lungs for healthy cases. This finding matches the clinical literature that lesion regions inside the lung are one of the significant risk factors for COVID-19 patients (\citeauthor{rajinikanth2020harmony} \citeyear{rajinikanth2020harmony}). Meanwhile, the infected branch also provides useful information by discovering the lungs' unnormal parts (colored in orange). While these lesions are rarely present or appear sparingly in healthy cases, it is clear that this feature plays an important factor in assessing the patient's condition. Finally, given data points distributed close to the margin separate the COVID-19 and non-COVID cases, learned heat-map regions overlapped for both lung and non-lung regions, indicating the uncertainty of the model's prediction. In such situations, utilizing other tests to validate results and the clinician's experience is a necessary factor in evaluating the patient's actual condition instead of just relying on the diagnosis of the model. 
For this property, we once again understand the importance of an explainable model. Without such information, we have a high risk of making mistakes using automated systems while we could not predict all possible situations.

\section{Conclusion}
In this paper, we have presented a novel approach to improve deep learning-based systems for COVID-19 diagnosis. Unlike previous works, we got inspired by considering radiologists' judgments when examining COVID-19 patients; thereby, relevant information such as infected regions or heat-maps of injury area is judged for the final decision. Extensive experiments showed that leveraging all visual cues yields improved performances of several baselines, including two best network architectures (ResNet50 and DenseNet169) from (\citeauthor{he2020sample} \citeyear{he2020sample}) and three other states of the arts from recent works. Last but not least, our learned features provide more transparency of the decision process to end-users by visualizing positions of attention map.  As effective treatments are developed, CT images may be combined with additional medically-relevant and transparent information sources. In future research, we will continue to investigate this in a large-scale study to improve the proposed system's performance towards explainability as an inherent property of the model.  

%
%

%
%



\bibliography{2021-AAAI-Covid}
\bibliographystyle{unsrt}  

\includepdf[pages=-]{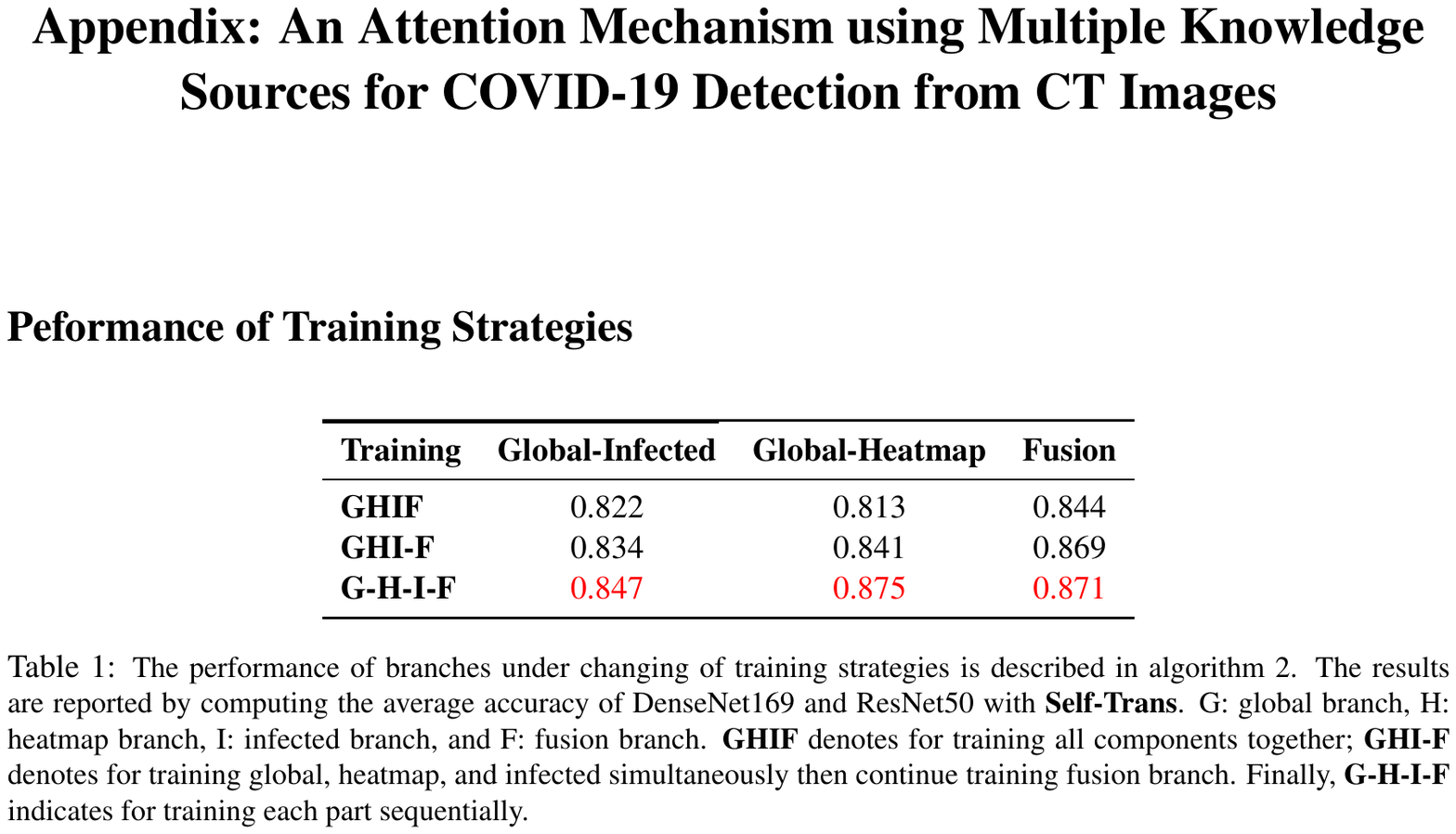}


\end{document}